\documentstyle[epsf]{article}

\font\tenrm=cmr10
\font\tenit=cmti10
\font\elevenbf=cmbx10 scaled\magstep 1
\font\elevenrm=cmr10 scaled\magstep 1
\font\elevenit=cmti10 scaled\magstep 1

\def\gtap{\raisebox{-.4ex}{\rlap{$\sim$}} \raisebox{.4ex}{$>$}}
\def\ltap{\raisebox{-.4ex}{\rlap{$\sim$}} \raisebox{.4ex}{$<$}}

\renewenvironment{thebibliography}[1]
 { \elevenrm
   \begin{list}{\arabic{enumi}.}
    {\usecounter{enumi} \setlength{\parsep}{0pt}
     \setlength{\itemsep}{3pt} \settowidth{\labelwidth}{#1.}
     \sloppy
    }}{\end{list}}

\begin{document}
\font\fortssbx=cmssbx10 scaled \magstep2
\hbox to \hsize{
\hfill\vbox{\hbox{\bf IUHET-310}
            \hbox{hep-ph/9508209}
            \hbox{July 1995}} }

\begin{center}
\vglue -0.3cm
{
 {\elevenbf        \vglue 10pt
     THE {\large \boldmath$t\overline{t}$} THRESHOLD AT A MUON
COLLIDER\footnote{
Presented at the Workshop on Particle Theory and
Phenomenology: Physics of the Top Quark, Iowa State University,
May 25-26, 1995.
}
\\}
\vglue 1.0cm
{\tenrm M.~S.~Berger\\}
\baselineskip=13pt
{\tenit Physics Department, Indiana University \\}
\baselineskip=12pt
{\tenit Bloomington, IN 47405, USA\\}}

\vglue 0.8cm
{\tenrm ABSTRACT}

\end{center}

\vglue 0.3cm
{\rightskip=3pc
 \leftskip=3pc
 \tenrm\baselineskip=12pt
 \noindent
The beam energy spread is a major issue in future attempts to study the
$t\overline{t}$ threshold at $e^+e^-$ colliders.
Muon colliders are expected to naturally have narrow band beams making
them an ideal place to study the excitation curve.
We present the parameter determinations that are possible
from measuring the total cross section near threshold at a $\mu ^+\mu ^-$
collider.}

\vglue 0.2in
\baselineskip=14pt
\elevenrm

\vglue 0.2cm
{\elevenbf\noindent 1. Introduction}
\vglue 0.2cm

The production of top quarks near threshold has been studied vigorously in
the last few years$^{1-15}$.
This process provides a test of perturbative QCD, as well as
an accurate measurement of the top
quark mass $m_t$ and the strong gauge coupling constant $\alpha _s$. Since
theoretical problems are for the most part under control, smaller effects
such as the top-Higgs Yukawa coupling and the top quark width might be probed.
These previous studies have
concentrated on the capabilities of $e^+e^-$ machines.

Recently attention has been devoted to the possibility
of a muon collider$^{16-21}$.
The impact on top quark threshold studies arises in
two ways: (1) there is decreased initial state radiation (ISR) at a muon
collider compared to an electron collider, and (2) the
intrinsic beam spread and beamstrahlung of a muon collider is expected to
be much smaller than at $e^+e^-$ machines. These effects are attributed
to the fact that the muon is much heavier than the electron, and result in
a threshold curve which is more steeply rising. This offers the possibility of
an improved measurement of the relevant parameters affecting the shape of
the cross section.
By the time a muon collider would actually be constructed,
the top mass will be measured to a few GeV and a relatively inexpensive
special-purpose ring could be constructed for optimizing the luminosity at
$\sqrt{s}=2m_t$. The utility of a narrow band beam has also been demonstrated
for $s$-channel Higgs studies$^{19-21}$.

\vglue 0.6cm
{\elevenbf\noindent 2. The Top Quark Threshold Region}
\vglue 0.2cm

Fadin and Khoze demonstrated that the top quark threshold cross section is
calculable since the large top
quark mass puts one in the perturbative regime of QCD,
and the large top quark width effectively screens nonperturbative effects
in the final state.
There are two (equivalent) ways to obtain the total cross section near
threshold: solving for a three-point Green's function$^4$ in either
coordinate or momentum
space. Here we solve the Schr\"{o}dinger's equation
\begin{eqnarray}
\left [-{{\Delta }\over m_t}+V(r)-\left (E+i{{\Gamma _{\Theta}}\over 2}\right )
\right ]G({\bf x};E)&=&\delta^3({\bf x})\;,
\end{eqnarray}
where $\Gamma _{\Theta}$ is the (running) toponium width,
and $E=\sqrt{s}-2m_t$.
The potential $V(r)$ is given for small $r$ by two-loop perturbative QCD and
for large $r$ by a fit to quarkonia spectra.
The total cross section is then proportional to ${\rm Im}\; G({\bf x}=0;E)$.
In addition to the usual Yukawa term $V_H(r)$ in
the potential at a
muon collider, there is an additional $s$-channel Higgs contribution to the
cross section since the muon has a larger Yukawa coupling than does the
electron. This contribution is much smaller than the usual photon and $Z$
exchanges considered here.

\vglue 0.6cm
{\elevenbf\noindent 3. Initial State Radiation}
\vglue 0.2cm

The larger mass of muon compared to the electron
implies that the initial state radiation in the
process $\mu ^+\mu ^-\rightarrow t\overline{t}$ is reduced compared to
$e ^+e ^-\rightarrow t\overline{t}$. Figure 1 shows the effects of including
initial state radiation (but not beam effects) for a top quark mass of
180 GeV. There is a reduction of the
cross section as well as a smearing out of the small resonance peak.

\begin{center}
\epsfxsize=3.9in
\hspace*{0in}
\epsffile{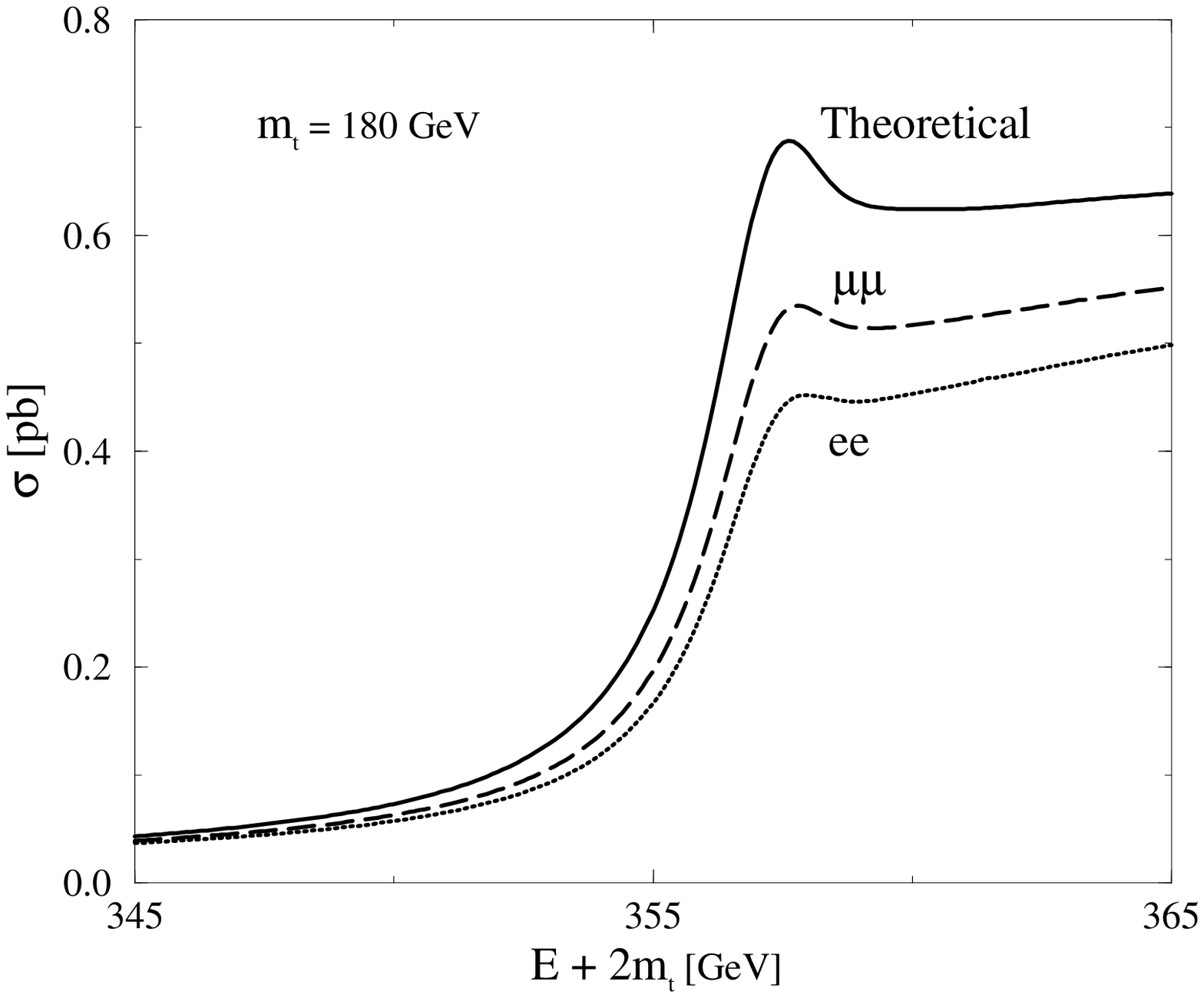}

\parbox{5.5in}{\small \small Fig.~1. Initial state radiation reduces the
total cross sections and smears the threshold region. The effect is smaller
for muons since they are heavier than electrons. The strong coupling is taken
to be $\alpha _s(M_Z)=0.12$.}
\end{center}

\vglue 0.6cm
{\elevenbf\noindent 4. Beam Effects}
\vglue 0.2cm

The beam energy spread is the major problem with precision measurements of the
top threshold region at an $e^+e^-$ collider. Ref.~\cite{bagliesi}
demonstrated the effects of beam smearing for some proposed machine designs,
and argued that a narrow beam was essential for studying the top quark
threshold region.
A muon collider should provide a naturally very narrow beam with rms deviation
$R\ltap 0.1\%$. The energy spectrum of each muon beam is expected to be roughly
Gaussian in shape, but for $t\overline{t}$ studies the exact shape is not
important if $R\ltap 0.1\%$. We take
$R=0.1\%$ here, but the results are not changed significantly if the resolution
is improved. We simulate the beam spread of the $e^+e^-$ machine by taking
$R=1\%$; significant detailed studies have taken more realistic parameters for
the electron collider$^{5,11}$,
and have included beamstrahlung effects.
The corresponding rms deviation $\sigma $ in $\sqrt{s}$ is given by
\begin{equation}
\sigma = (0.25~{\rm GeV})\left({R\over 0.1\%}\right)\left({\sqrt s\over {\rm
360\ GeV}}\right) \;,
\end{equation}
which for a muon collider (with $R\ltap 0.1\%$) is of the same
order as the measurement one hopes to make in the top mass.
For large values of R ($\gtap 0.5\%$), the shape of the cross section in the
threshold region depends on the beam profile; we assume here that the $e^+e^-$
collider has a Gaussian profile for simplicity. A high resolution
determination of the collider energy profile would be desirable to deconvolute
the smearing of the threshold curve. For a muon collider on the other hand, no
such measurement of the beam profile would be necessary.
Figure 2 shows the effects of including beam smearing of the threshold curve.

\begin{center}
\epsfxsize=3.9in
\hspace*{0in}
\epsffile{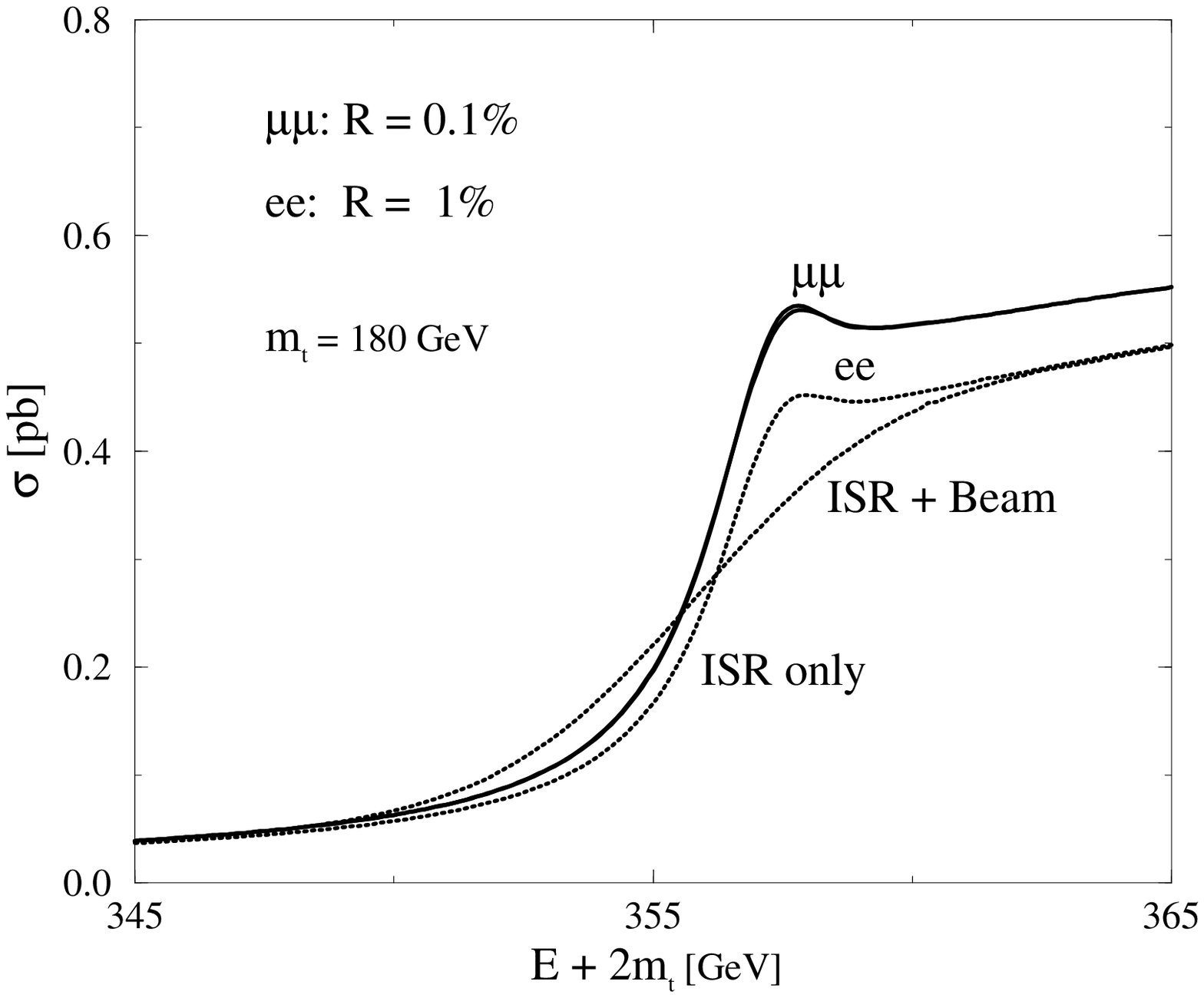}

\parbox{5.5in}{\small \small Fig.~2. The threshold curves are shown for
$\mu ^+\mu ^-$ and $e^+e^-$ machines including ISR and with and without
beam smearing. Beam smearing has only a small effect
at a muon collider, whereas at an electron collider the threshold region is
significantly smeared. The strong coupling is taken
to be $\alpha _s(M_Z)=0.12$.}
\end{center}

\vglue 0.6cm
{\elevenbf\noindent 5. Parameter Determination}
\vglue 0.2cm

The $t\overline{t}$ threshold shape is a sensitive function of the strong
coupling constant and top quark
mass as shown in Figs. 3 and 4 respectively.

\begin{center}
\epsfxsize=3.8in
\hspace*{0in}
\epsffile{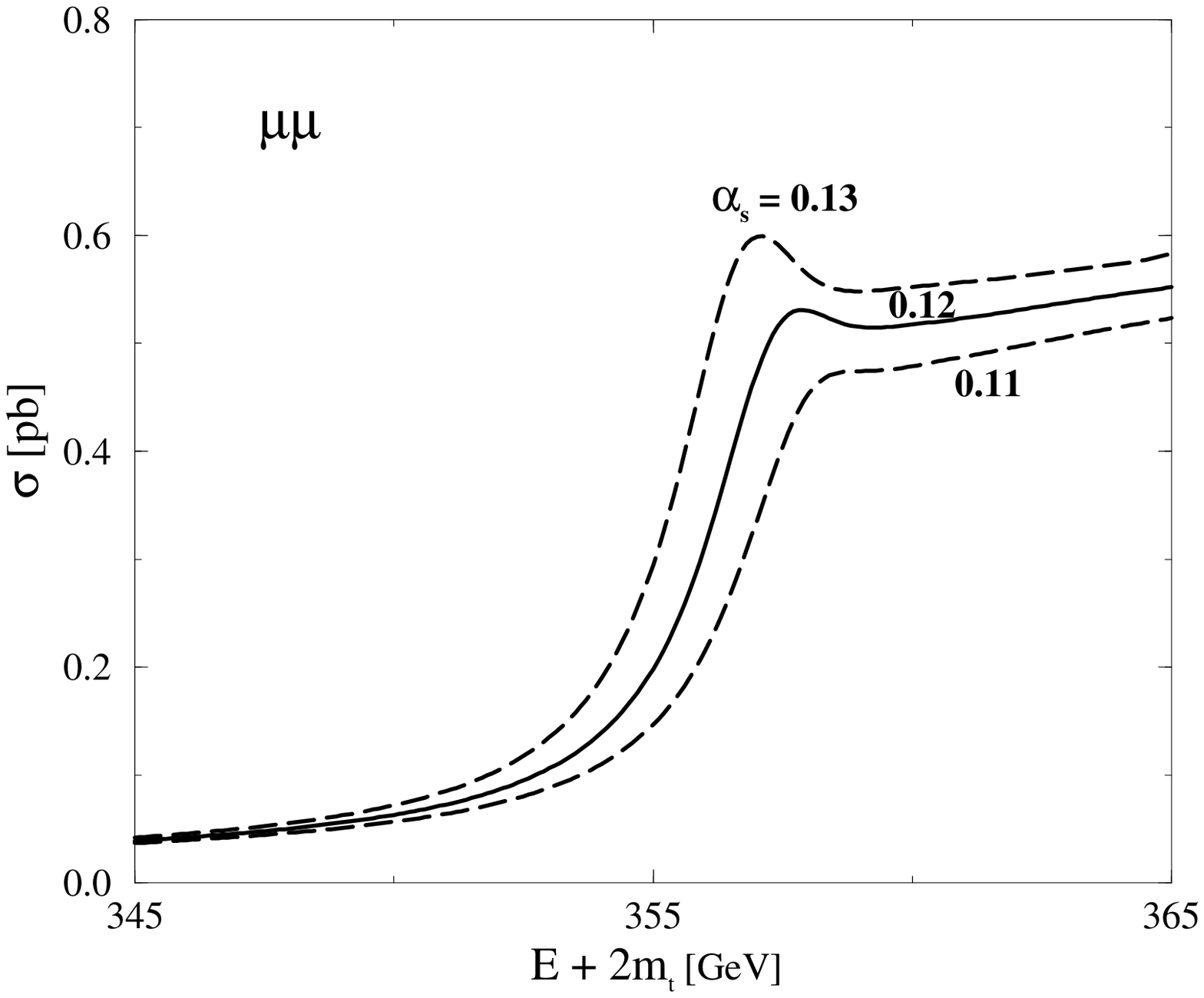}

\parbox{5.5in}{\small \small Fig.~3. Changing the value of the strong coupling
constant $\alpha _s(M_Z)$ influences
the threshold region. Large values lead to
tighter binding and the peak shifts to lower values of $\sqrt{s}=E+2m_t$. Both
ISR and beam smearing have been included, and the top quark mass is taken to
be 180 GeV.}
\end{center}

\begin{center}
\epsfxsize=3.8in
\hspace*{0in}
\epsffile{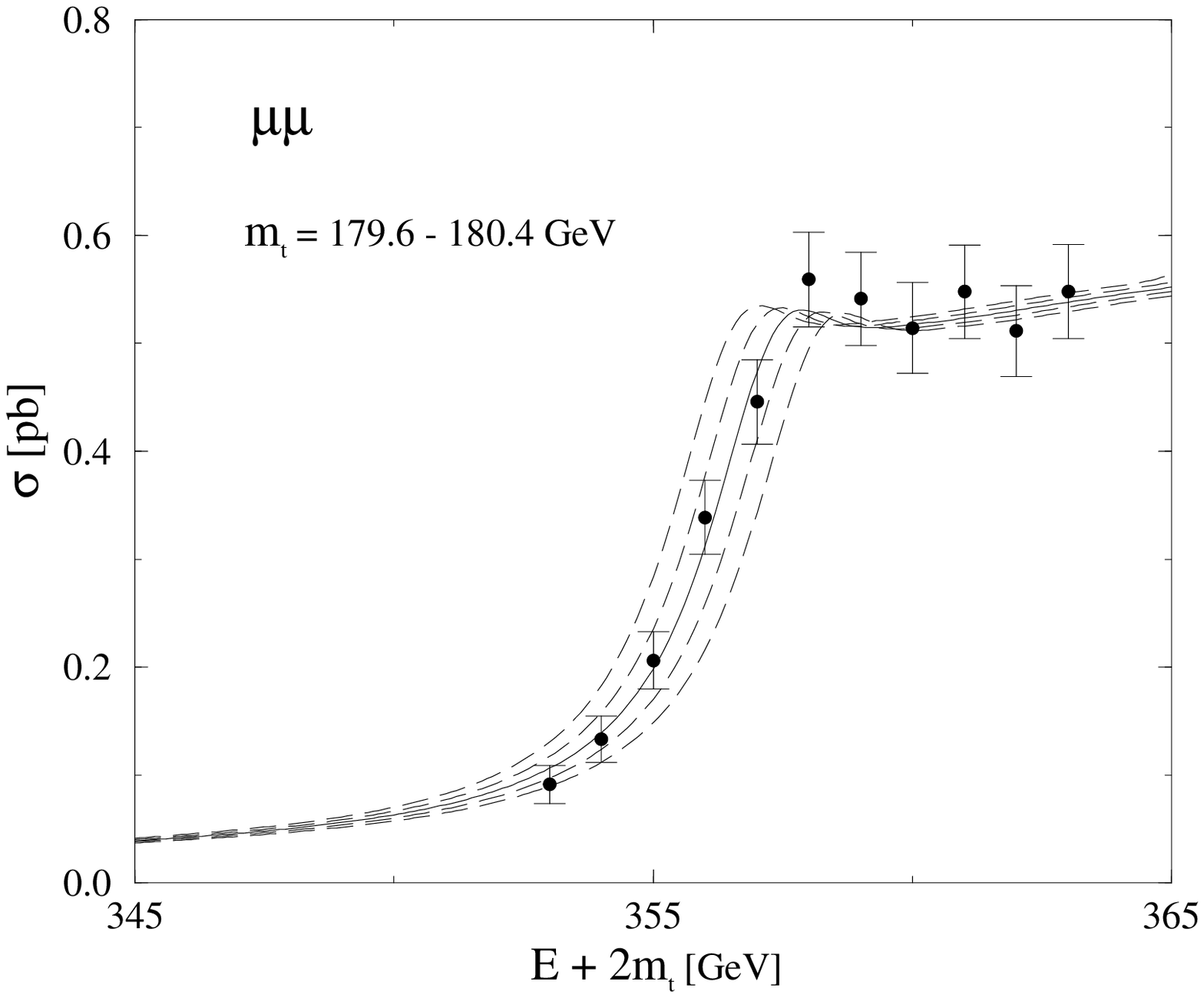}

\parbox{5.5in}{\small \small Fig.~4. Sample data obtained assuming a scan over
the threshold region devoting 1~fb$^{-1}$ luminosity to each data point.
A detection efficiency of 29\% has been assumed in obtaining
the error bars. The threshold curves correspond to $m_t$ in $0.2$~GeV
increments. Both
ISR and beam smearing have been included, and the strong coupling
$\alpha_s(M_Z)$ is taken to be 0.12.}
\end{center}

Decreasing $m_t$ and increasing $\alpha _s(M_Z)$ have similar effects on the
threshold curve. To assess the precision of parameter determinations from
measuring the total cross section, we
take some hypothetical sample data as shown in Fig.~4, generated assuming
1~fb$^{-1}$ integrated luminosity is used to measure the cross section in
1~GeV intervals and assuming the
nominal values $m_t=180$ GeV and $\alpha _s(M_Z)=0.12$. Following
Ref.~\cite{fms} we assume a 29\% detection efficiency to isolate the
$W\to qq$ final state from background.
The likelihood fit to the two parameters $m_t$ and $\alpha _s(M_Z)$ is shown
as the $\Delta \chi ^2$ contour plot in Fig.~5.
The outer curve is the $90\%$ confidence level ($\Delta \chi ^2=4.6$) for
$m_t$ and $\alpha _s$ jointly, while the inner curve is the $1\sigma$
confidence level ($\Delta \chi ^2=2.3$).

\begin{center}
\epsfxsize=3.9in
\hspace*{0in}
\epsffile{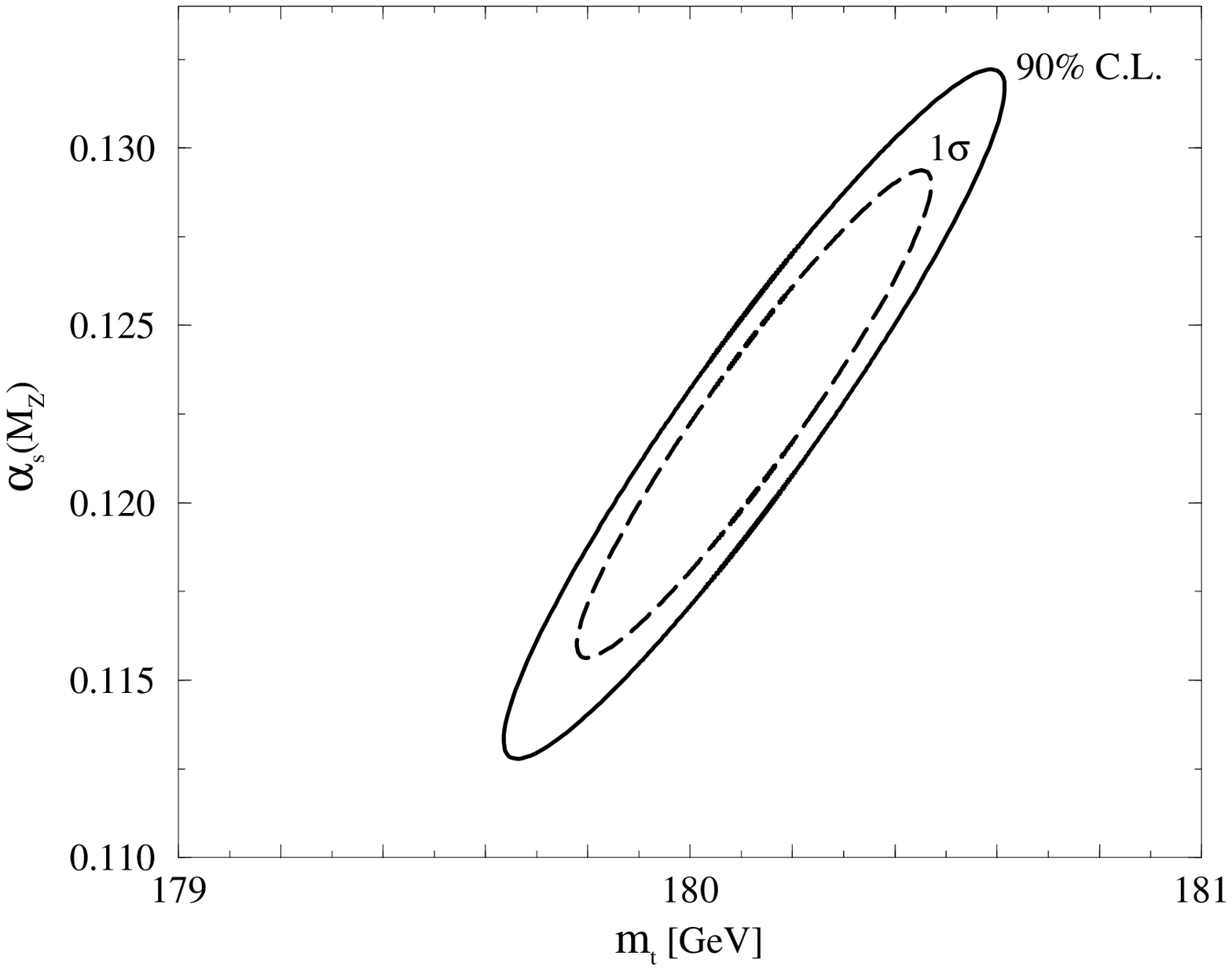}

\parbox{5.5in}{\small \small Fig.~5. The $1\sigma$ and $90\%$ confidence
limits for the sample data shown in Fig.~4.}
\end{center}

A top quark mass around 180 GeV can be measured to about 300 MeV at
$90\%$ confidence level (the $\Delta \chi ^2=2.7$ contour is not shown).
This is roughly a factor of two better than the same measurement at an
$e^+e^-$ machine when realistic beam effects are included$^{11}$. The
improvement in the measurement of $\alpha _s(M_Z)$, however, is more modest.
As at an $e^+e^-$ collider,
one will be able to
make use of the top momentum measurement to improve on the parameter
determinations from the total cross section$^{7,9,11}$.
A more complete treatment of the $t\overline{t}$ threshold measurement at
a muon collider will be presented in Ref.~\cite{bbgh2}.

\vglue 0.6cm
{\elevenbf\noindent 6. Acknowledgements}
\vglue 0.2cm
This research was supported
in part by the U.S.~Department of
Energy under contract no.~DE-FG02-91ER40661.

\newpage
{\elevenbf\noindent 7. References}
\vglue 0.2cm

\end{document}